\documentclass[fleqn,usenatbib]{mnras}

\usepackage{newtxtext,newtxmath}

\usepackage[T1]{fontenc}

\DeclareRobustCommand{\VAN}[3]{#2}
\let\VANthebibliography\thebibliography
\def\thebibliography{\DeclareRobustCommand{\VAN}[3]{##3}\VANthebibliography}

\usepackage{ragged2e}
\usepackage{color}

\usepackage{graphicx}	
\usepackage{amssymb}
\usepackage{amsmath}	

\usepackage{tikz}
\usepackage{epstopdf}

\newcommand{\fu}{~erg~s$^{-1}$~cm$^{-2}$}

\def\swtr{Swift~J1727.8$-$1613}

\newcommand{\art}{ART-XC}
\newcommand{\srg}{{\it SRG}}
\newcommand{\integral}{{\it INTEGRAL}}
\newcommand{\swift}{{\it Swift}}

\def\arcsec{\hbox{$^{\prime\prime}$}}

\usepackage[normalem]{ulem}

\long\def\***#1{\textcolor{blue}{\textbf{\sffamily ***#1***}}}

\title[Early stages of Swift\,J1727.8-1613 outburst]{Hard X-rays and QPO in Swift\,J1727.8-1613: the rise and plateau of the 2023 outburst}

\author[I. Mereminskiy et al.]{%
I. Mereminskiy,$^{1}$\thanks{E-mail: i.a.mereminskiy@gmail.com}  A. Lutovinov,$^{1}$, S. Molkov,$^{1}$ R. Krivonos,$^{1}$ A. Semena,$^{1}$  S. Sazonov,$^{1}$ A. Tkachenko$^{1}$ \newauthor and R. Sunyaev$^{1,2}$
\vspace*{-0.5\baselineskip}\\
$^{1}$Space Research Institute RAS, 84/32 Profsoyuznaya str, Moscow 117997, Russia\\
$^{2}$Max Planck Institute for Astrophysics, Karl-Schwarzschild-Str. 1, 1317, D-85741 Garching, Germany\\
}

\date{Accepted XXX. Received YYY; in original form ZZZ}

\pubyear{2023}
\begin{document}
\label{firstpage}
\pagerange{\pageref{firstpage}--\pageref{lastpage}}
\maketitle

\begin{abstract}
We report on the detection of type-C quasi-periodic oscillations during the initial stages of the outburst of Swift J1727.8--1613 in 2023. Using data of the \integral\ observatory along with the data of the \srg/\art\ and \swift/XRT telescopes the fast growth of the QPO frequency was traced. We present a hard X-ray lightcurve that covers the initial stages of the 2023 outburst -- the fast rise and plateau -- and demonstrate that the QPO frequency was stable during the plateau. The switching from type-C to type-B QPO was detected with the beginning of the source flaring activity. We have constructed a broad-band spectrum of \swtr\ and found an additional hard power-law spectral component extending at least up to 400 keV. Finally, we have obtained an upper limit on the hard X-ray flux at the beginning of the optical outburst and estimated the delay of the X-ray outburst with respect to the optical one.
\end{abstract}

\section{Introduction} 
\label{sec:intro}

Low-frequency quasi-periodic oscillations \citep[LF QPO, see ][for a review]{wijnands99,ingram19_qpo_rev} of the X-ray flux are commonly observed during different stages of outbursts of low-mass X-ray binaries (LMXB). Usually, the first to emerge are type-C QPOs \citep{casella05}, which appear during the outburst rise phase, while the source is in a low/hard state \citep{tanaka96,remillard06}. Despite significant theoretical efforts \citep[see, e.g.,][for one of earlier predictions of QPO in black hole systems]{1973SvA....16..941S}, there is no model that could reproduce the entire plethora of observed phenomena. Different models attribute the observed variability to Lense-Thirring precession of the central parts of the accretion flow \citep{stella98,ingram09}, oscillations in the corona \citep{cabanac10}, coupling between the accretion disk and the Comptonizing corona \citep{mastichiadis22}, etc. \citep[see][for discussion of other models]{ingram19_qpo_rev}.  

Although it is interesting to trace the QPO evolution all the way to the start of the outburst, that is rarely possible given the unpredictable nature of LMXB X-ray outbursts \citep[although sometimes optical monitoring could warn us about an upcoming outburst,][]{russell19_leadingOIR}, limited sensitivity of all-sky X-ray monitors, and the shortness of the rise stage, which  usually lasts only few days. 

A new X-ray transient, dubbed \swtr, was detected in hard X-rays on Aug 24, 2023 by both \swift/BAT \citep{page23GCN34537} and \integral\, \citep[IBAS weak trigger 10373/0]{IBAS}.  Later on, a number of optical \citep{atlas23,2023ATel16225....1B} and X-ray telescopes \citep{2023ATel16205....1N,2023ATel16207....1O,2023ATel16210....1L, 2023ATel16217....1S, 2023ATel16242....1D} also observed it. Optical spectroscopy \citep{2023ATel16208....1C} revealed the presence of bright emission lines of hydrogen and helium, leading to the classification of the source as a LMXB candidate. X-ray spectral and temporal \citep{2023ATel16215....1P, 2023ATel16219....1D} characteristics of the source were also typical for LMXB outbursts.  

In this paper, we report on the detection of type-C QPO in the earliest stages of the outburst of \swtr\ with different X-ray telescopes. We were able to trace the evolution of the QPO frequency in detail during the first two weeks of the outburst, while the source transitioned from rapid rise to plateau. We also obtained and characterized the hard X-ray spectrum of the source in the broad energy range of 5--500 keV during the plateau phase of the outburst. \swtr\, is confidently detected up to 400 keV.

Because the wide-field telescopes aboard the \integral\, observatory are rarely used for aperiodic timing studies, we provide the public with an open dataset, which could be used together with other multiwavelength observations in order to test models of LF QPO generation.
\begin{figure*}
    \centering
    \includegraphics[width=0.95\textwidth]{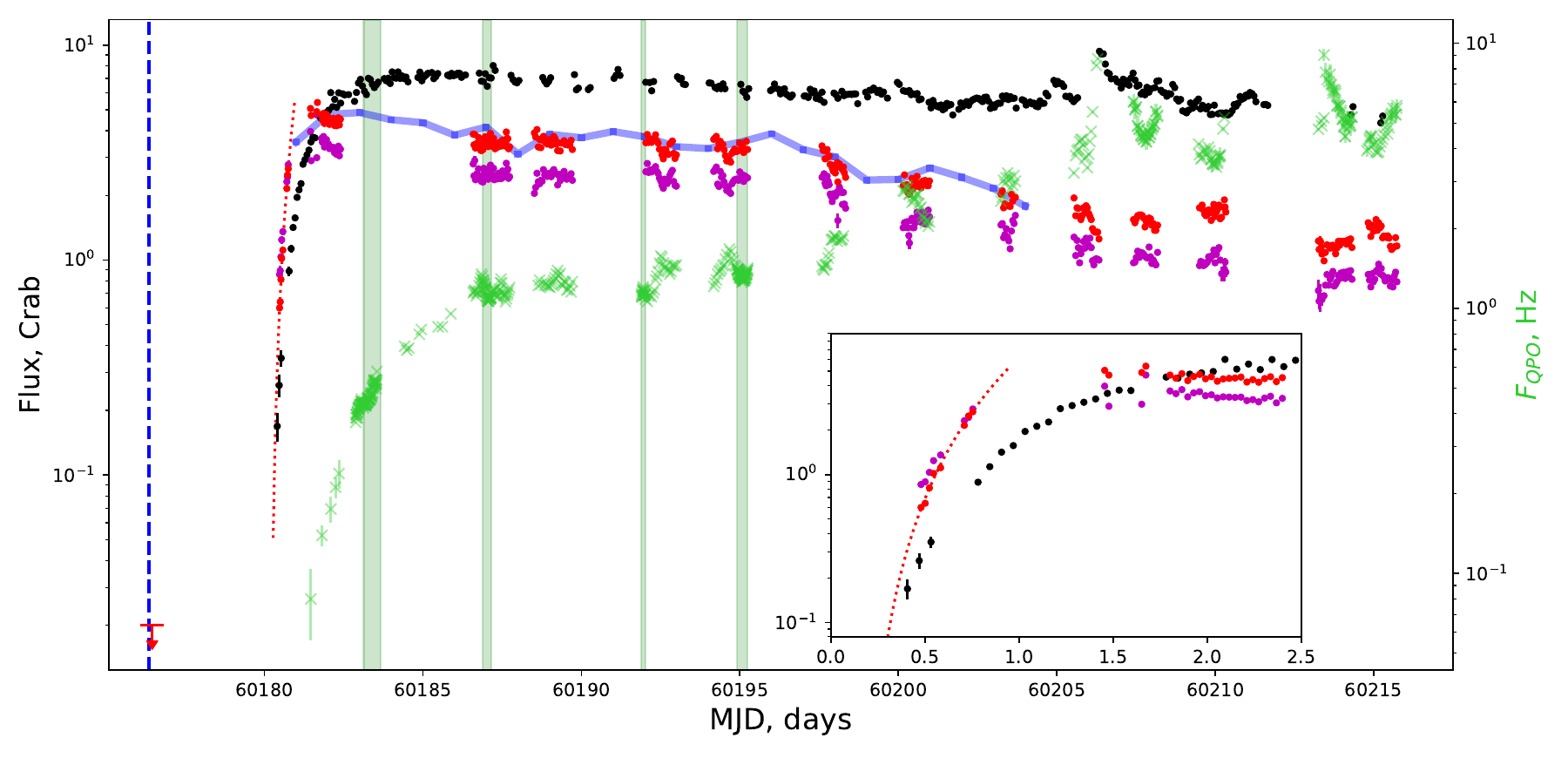}
    \caption{Long-term X-ray lightcurve of \swtr. Black points show MAXI 2--20 keV fluxes and red (magenta) markers show IBIS/ISGRI measurements in the 30--80 keV (80--150 keV) band, the upper limit corresponds to the non-detection at the earliest stages of the outburst. The blue solid line shows the \swift/BAT flux measurement in 15--50 keV, arbitrary rescaled to match the IBIS data points. The thin red dotted line shows a power-law fit to the first 8 \integral\, points. The vertical blue dashed line shows the first detection of optical emission by {\it ATLAS}. The green areas indicate the periods of \art\, observations. The inset shows a zoom on the rise phase, the X-axis reads days since MJD~60180.
    Green crosses (right axis) shows the QPO frequency evolution.}
    \label{fig:olc}
\end{figure*}

\section{Data analysis}
\label{sec:da}

To better characterize the early evolution of the source flux and to study its temporal properties, we used data from a number of X-ray telescopes. To construct a continuous long-term X-ray lightcurve of the outburst, we used data from MAXI \citep{matsuoka09}, covering the softer part of spectrum (2--20 keV). Because the MAXI lightcurve is contaminated by the nearby bright Galactic source GX\,9+9, we decided not to use the MAXI data before MJD~60180.4, i.e. a few hours before the first hard X-ray detection.

We produced a long-term hard X-ray lightcurve using numerous serendipitous and target-of-opportunity \integral\, observations (see details in Sec.\ref{sec:integral}). To fill the gaps between adjacent \integral\ revolutions, we also used data from the \swift/BAT \citep{sbat05} transient monitor \citep{krimm13}. Since the energy bands of the two instruments are slightly different, IBIS (30--80 keV) vs BAT (15--50 keV), we rescaled the BAT fluxes to match the IBIS measurements. 

A number of observations of \swtr\ during the first days of the outburst were performed by the \swift/XRT telescope \citep{sxrt05}. These data are crucial for us, since they enable a timing analysis. We selected observations performed in windowed timing mode (ObsIDs: 1186959005-006) to produce 0.5--10 keV lightcurves for each snapshot.

\subsection{\integral\ observations}
\label{sec:integral}

Thanks to the large field of view of the IBIS coded-mask telescope \citep{2003A&A...411L.131U} on board the {\integral} observatory \citep{2003A&A...411L...1W}, the sky region of the X-ray transient {\swtr} was covered during observations of the Galactic Center region shortly before the outburst, during the rapid increase of the flux and during the outburst itself. In this work, we use the data acquired by the ISGRI low-energy detector layer \citep{2003A&A...411L.141L} of the IBIS telescope. For the timing analysis we also use data from the JEM-X telescope on board \integral, which is sensitive in the 3--35 keV energy range \citep{2003A&A...411L.231L}. 

We reduced the IBIS/ISGRI data with the \integral\ data analysis software developed at IKI \citep[see e.g.,][and references therein]{2010A&A...519A.107K,2012AandA...545A..27K,2014Natur.512..406C}, applying  the latest energy calibration for the registered IBIS/ISGRI detector events with the INTEGRAL Offline Scientific Analysis (OSA) version 11.2 provided by the INTEGRAL Science Data Centre (ISDC) Data Centre for Astrophysics \citep[][]{2003A&A...411L..53C}. 

We produced a sky image for every individual \integral\ observation  (referred to as a {\it Science Window}, or {\it ScW}). The flux scale in each {\it ScW} sky image was adjusted to the flux of the Crab Nebula measured in all available observations taken in 2023. Due to the loss of sensitivity at low energies caused by the gradual long-term degradation of the ISGRI detector, we set a low limit of 25~keV for our analysis. For spectral analysis, we generated images in 30 energy intervals logarithmically spaced between 25 and 450 keV.

For a fast timing analysis and the analysis of JEM-X data, we used the INTEGRAL Offline Scientific Analysis version 11.2. Before constructing the ISGRI light curves, we first recalculated the photon energies in the instrument's event files using up-to-date calibration data. To this end, we launched the \textsc{OSA} procedure \texttt{ibis\_science\_analysis} to the COR level. Next, using the \texttt{ii\_pif} procedure, we calculated the ``open part'' of each pixel (i.e. pixel-illumination-fraction, PIF) of the detector for the direction to the source. Then, we constructed a light curve in a given energy band with a 0.01~s time resolution for ``fully open'' pixels subtracted from it a light curve obtained from the ``fully closed'' pixels with the same time resolution, taking into account the number of pixels used for each light curve. In our case, when there is only one bright source in the field of view of the IBIS telescope, this is the simplest and most correct way to reconstruct a light curve.

\subsection{\srg/\art\ observations}
\label{sec:art}
Shortly after the discovery of \swtr, we observed it with the Mikhail Pavlinsky ART-XC telescope \citep{pavlinsky21} on board the \srg\ observatory \citep{SRG21}. The first observation, conducted on Aug 27, 2023, caught the source during the rise stage. Three later observations, on Aug 31, Sept 5 and 8, 2023, fell on the plateau phase on the lightcurve. \art\ data were processed using the {\sc artproducts} v1.0 software with the latest CALDB {\sc v20220908}. Spectra and light curves were extracted from a circular region  of radius $R = 225\arcsec$ centered at the source position. For the extraction of the light curve, a wide energy range of 4--25 keV was used. The spectral analysis was performed in a slightly different energy band of 5--25 keV, where the spectral response of the telescope is known better.

\section{Outburst lightcurve and early optical observations}
\label{sec:olc}

Figure~\ref{fig:olc} shows the X-ray lightcurve of the outburst based on MAXI (2--20 keV) and IBIS (30--80 and 80--150 keV) data. The first detection of the source in these bands happened at a flux level of $\approx0.1$~Crab. In hard X-rays, the flux peaked approximately at MJD~60181.6 and then started to decline. The 2--20 keV lightcurve shows no obvious peak. After the initial rise, the source reached a plateau of $\approx7$~Crab and stayed at this level for about a week. Afterwards, the plateau changed to a modest decline, seen also in the hard band. Later on, the source started to exhibit soft X-ray flares, indicating a transition to a disk dominated soft-intermediate state. 
The overall lightcurve is typical of LMXBs in outburst, before transition to the soft state.

In recent years, monitoring observations of the optical sky have become widespread, with several projects covering significant parts of the sky with daily cadence. In conjunction with space-based X-ray observatories, this makes it possible to trace the evolution of outbursts of newly discovered X-ray transients simultaneously in the optical and X-ray bands. Together with dedicated programs (e.g. XB-NEWS, \citealt{russell19_leadingOIR}), this monitoring shows that optical emission often precedes X-ray outburst.   

The optical counterpart of \swtr\, was detected by the ATLAS project \citep{tonry_18} as early as MJD~60176.368 \citep{atlas23} during the rapid rise. This brightening was bracketed by serendipitous \integral\, observations taken between MJD~60176.095 and MJD~60176.626. This allowed us to obtain a strict $4\sigma$ upper limit of 18~mCrab (or $3.5\times10^{-11}$ erg cm$^{-2}$ s$^{-1}$) on the average 30--60 keV flux of the source during this period. The optical flux at the same epoch was $\approx$100 $\mu$Jy in the ATLAS $o$-filter. Using the value of the neutral hydrogen column density in the direction of \swtr,  $N_{\mathrm{H}}$ = $4\times10^{21}$ cm$^{-2}$, determined from X-ray data \citep{2023ATel16219....1D} and adopting $A_{V}/N_{H} = 0.48$ \citep{guver09}, we estimate the total extinction in the $o$-filter at $\approx 1.4$ magnitudes. Hence, the monochromatic optical flux from \swtr\, was $\approx 2\times10^{-12}$ erg cm$^{-2}$ s$^{-1}$. We can thus place a conservative upper limit on the X-ray to optical luminosity ratio: $L_{\rm X}/L_{\rm opt} \lesssim 20$. 

Both the MAXI and \integral\, data show that the rapid brightening of the source started shortly before MJD~60180.5. 
To determine the epoch of onset of the X-ray outburst more precisely, we fitted a power law to the IBIS measurements of the flux on the rise (the first 8 observations) and obtained $F_{X} \propto (t-60180.2)^{2.4}$. Therefore, the X-ray outburst started with a delay of at least 3.8 days with respect to the optical one.

Such an X-ray-to-optical lag, previously observed in many  LMXBs \citep[e.g.][]{orosz97,jain01,wren01,buxton04}, can be explained in the disk-instability model \citep[see Sec. 5.1 in ][]{DHL01}. Using the scaling relation from \citet{bernardini16V404}, we can estimate the radius $R_{o}$ at which the optical outburst started. Assuming the midplane disk temperature to be $T = 40 000 $ K, adopting the viscosity parameter $\alpha = 0.2$ and assuming that the X-ray outburst starts when the inner radius of the thin disk reaches $R_{X} = 5\times 10^{8}$ cm \citep[as in ][]{DHL01}, we get $R_{o} \approx 10^{9}$ cm or about 700$R_{g}$ for a $10\,M_{\odot}$ black hole.
 
\section{Timing features}
\label{sec:QPOs}
Long hard X-ray observations by \integral\, are well suited to search for LF QPOs in the initial stages of bright LMXB outbursts \citep[e.g.][]{mereminskiy18}. We examined all available data from the JEM-X and IBIS telescopes in order to search for prominent features in the power spectra of \swtr. We also searched through \art\, and Swift/XRT data in order to get as many measurements of the QPO fundamental frequency as possible. We fitted each power spectrum (in Leahy normalization, 0.01--50 Hz) by a model consisting of four components: a constant for Poisson noise, a wide zero-centered Lorentzian for low-frequency noise and two narrow Lorentzians with tied frequencies to describe the fundamental QPO and its harmonic. We provide all reliable QPO measurements in Tab.~\ref{tab:qpos} along with Fig.~\ref{fig:olc}. 
During the flaring activity after MJD~60205, the QPO switched to the type-B with values up to $\sim10$ Hz (see Fig.\,\ref{fig:olc}). Since these QPO are beyond the scope of this paper we excluded them from analysis.

\begin{figure}
    \centering
    \includegraphics[width=0.975\columnwidth]{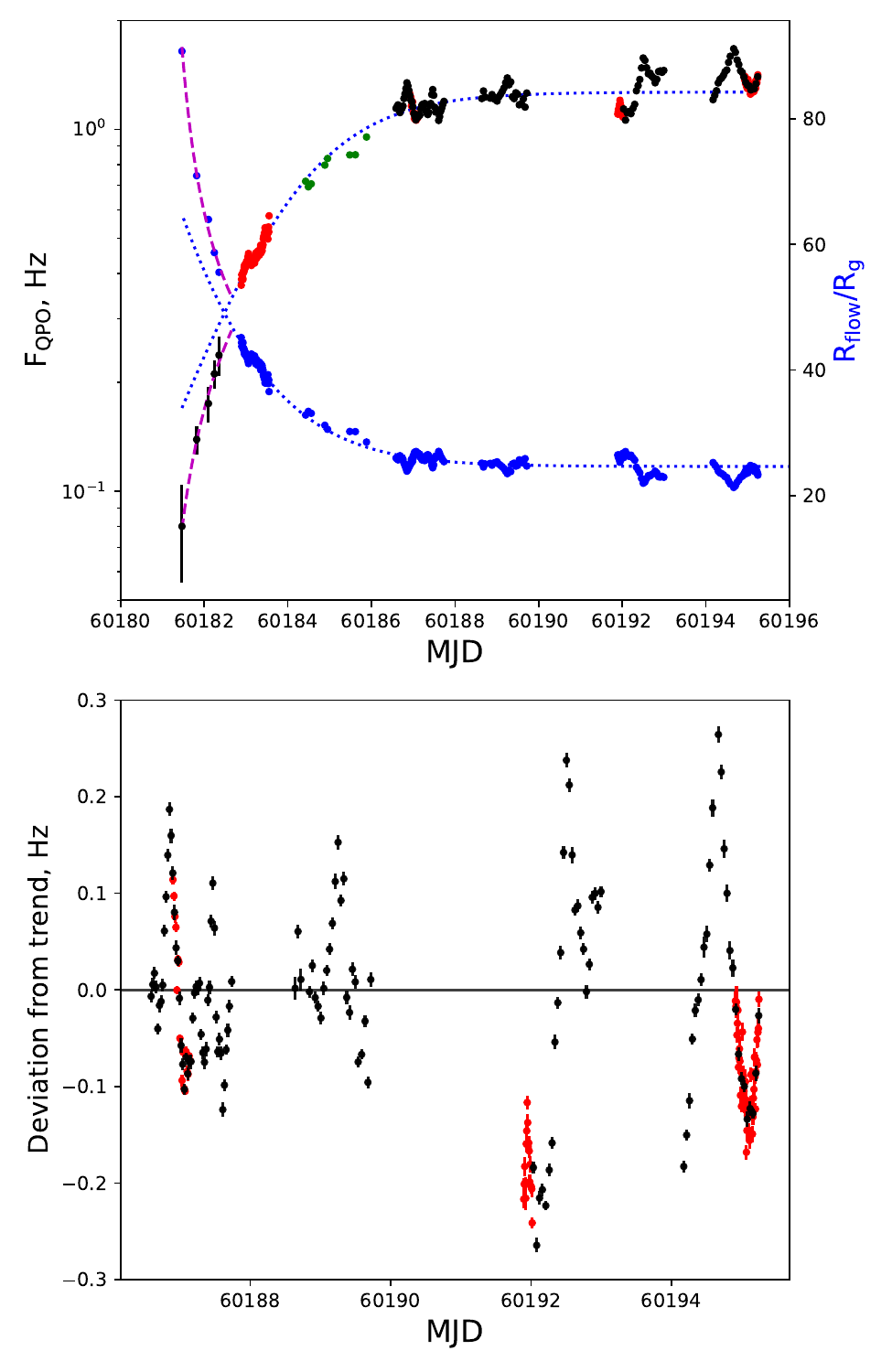}
    \caption{{\it Upper panel:} Evolution of the QPO frequency (left axis), based on measurements with \integral\ (black dots), \art\ (red) and \swift/XRT (green), and of the inferred radius of the precessing hot flow (right axis). The dashed magenta line shows the best-fit linear approximation for the earliest measurements, and the blue dotted line shows the logistic model that describes the later plateau phase. 
    {\it Lower panel:} Deviation of the QPO frequency from a linear trend at the plateau stage.}
    \label{fig:qpo_r}
\end{figure}

The first secure detection of type-C QPO in IBIS was at $F_{QPO}\approx 80 $ mHz on MJD~60181.46, one day after the first detection of the source. Then QPOs were detected at frequencies $F_{QPO} = 140...240$ mHz on several occasions with JEM-X, when the source was in its field of view. It should be noted that these observations were part of a regular Galactic Center region survey (PI: Sunyaev), therefore they were not optimized for studying \swtr. Later on, the QPO frequency continued to grow until it reached a plateau of 1.2 Hz a week after the start of the outburst. This plateau lasted for another week. Fig.~\ref{fig:qpo_r} summarizes the early evolution of the QPO frequency. The initial frequency growth is nearly linear, whereas the later evolution can be roughly described by a logistic function with a characteristic width of 0.8 days. After the plateau frequency continued to grow, while the hard flux receded. 

It is interesting to see how the QPO frequency behaves during the plateau stage. Since both the soft and hard X-ray fluxes are nearly constant during this period, we can assume that the accretion disk was stable so that the QPO frequency drifted around some central value. After subtraction of the linear trend (estimated from the IBIS data alone, $\approx0.03$ Hz day$^{-1}$), it can be clearly seen (Fig.~\ref{fig:qpo_r}, lower panel) that the frequency shows little variability on short timescales (as evident from the dense \art\, measurements) while oscillating around the mean value on a timescale of $\approx10$ hours with a standard deviation of $0.09$ Hz.

Using equation (2) from \citet{ingram09} and assuming that $\mathrm{M_{BH}}=10, a=0, \zeta=0$ and the inner radius of the precessing flow depends on the black hole spin \citep{lubow02}, we can infer the outer radius of the hot flow and trace its evolution with time. We thus find that during the initial fast-rise phase, the precessing flow shrinks rapidly at a rate of $\approx 40\,\mathrm{R_{g}}\,\mathrm{day^{-1}} \approx 700$\,cm\,s$^{-1}$. At the plateau stage, the characteristic peak-to-peak speeds are lower: $15 \,\mathrm{R_{g}}\,\mathrm{day^{-1}}$.  

\section{Spectrum and hard X-ray emission}
\label{sec:spectrum}

A detailed spectral analysis of the \swtr\ X-ray emission at different phases of the outburst is beyond the scope of this paper and will be carried out elsewhere, using data from all \integral\ instruments, including the SPI telescope. Below, we only present the results obtained by the \art\ telescope and the \integral\ observatory during the plateau phase in order to characterize the general shape of the spectrum and to search for hard X-ray-- soft gamma-ray radiation.

\begin{figure}
    \centering
    \includegraphics[width=0.95\columnwidth]{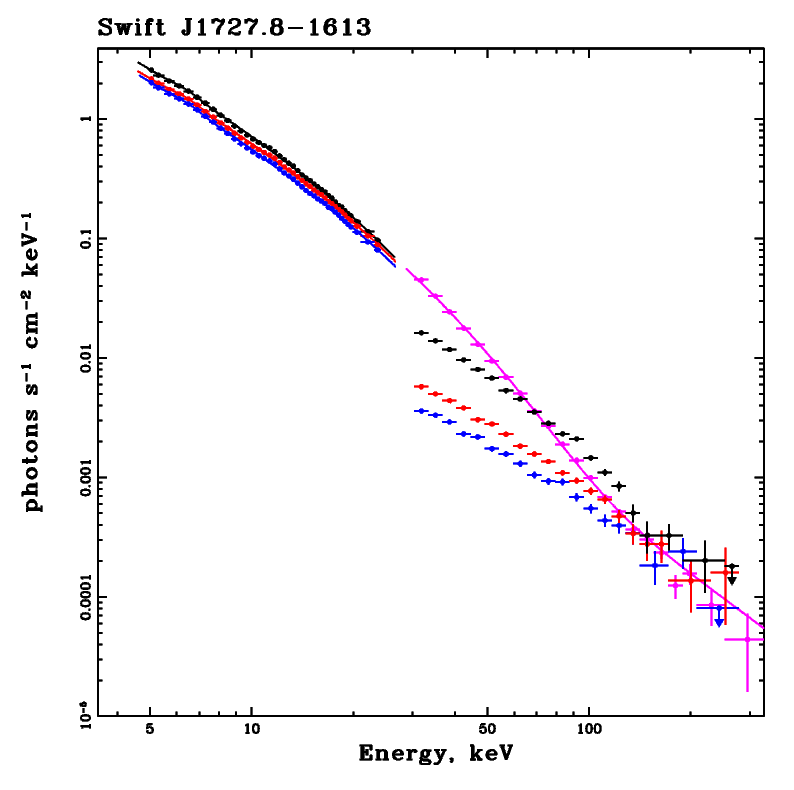}
    \caption{Broad-band energy spectrum of \swtr\ obtained with \srg/ART-XC (black, blue and red points in the 5-25 keV range) and \integral/IBIS (magenta) at three epochs during the plateau phase of the outburst. The \swtr\ hard X-ray spectra obtained during the outburst rise are shown by blue, red and black points in the 30-300 keV range. 
    }
    \label{fig:spec}
\end{figure}

The source was observed four times by the \art\ telescope in the period from Aug 27 to Sept 8, 2023. For spectral analysis we used data from second, third and fourth observations, which were conducted during the plateau phase. All three spectra have approximately the same shape (see Fig.\,\ref{fig:spec}) and can be well described by a power law with a slope of $\sim1.5$ and an exponential cutoff with an energy of 19--23 keV ({\sc cutoffpl} model in the {\sc XSPEC} package \citealt{1996ASPC..101...17A}). The spectrum also suggests the presence of a weak, broad emission line of neutral iron at 6.4 keV with an equivalent width of $\sim80$ eV \citep[see also][]{2023ATel16207....1O,2023ATel16219....1D}. The 5--25 keV flux in the \art\  observations evolved from $\sim1.7\times10^{-7}$ to $\sim1.4\times10^{-7}$ \fu. 

Taking the approximately constant spectral shape at the plateau phase into account, we next examined the broadband spectrum of the source in the energy range from 5 to several hundred keV to search for a hard X-ray tail. To this end, we used the aforementioned data from the \art\ telescope and data from simultaneous observations by the IBIS/\integral\ telescope. The resulting spectra are shown in Fig.\,\ref{fig:spec}. The data from both instruments are in good agreement with each other. This allows us to jointly approximate them by a combination of the {\sc cutoffpl} model described above and an additional power-law component. The latter component is needed to describe the high-energy radiation and has a slope of $\sim2.1$.

As mentioned above, we use a simple phenomenological model only for a general description of the shape of the spectrum and the search for hard radiation. More complex models, including those that taking into account Comptonization, contribution of the accretion disk, the reflected component, etc. will be applied at the next step.

From Fig.\,\ref{fig:spec} it can be seen that even on a limited data set, hard X-ray emission from \swtr\ is significantly detected up to $\simeq230$ keV. This, as well as the presence of a power-law hard X-ray tail in the spectrum, suggests the possibility of detecting even more energetic photons from the source. To this end, we used all available IBIS/\integral\ data obtained in the period from Aug 24 to Sept 26 and constructed images of the sky area around \swtr\ at energies above 250 keV. We found that \swtr\ was significantly registered up to energies $\simeq400$ keV. In particular, the source flux in the 350--400 keV energy range is $0.60\pm0.13$\,Crab. Yet harder radiation was not detected from the source, and the corresponding upper limit is 200 mCrab ($1\sigma$) in the 400--500 keV range. 

Finally, using the IBIS data we can trace the evolution of \swtr\ hard X-ray spectrum during the outburst rise (Fig.\,\ref{fig:olc}). To this end, we divided this phase into three parts of $\simeq5-6$ ks each and constructed a spectrum for each of them. Approximation by the {\sc cutoffpl} model shows that the photon index is  approximately the same, 1.25--1.35, for all three spectra, while the cutoff energy decreases from $\sim150$ keV at the very beginning of the outburst (blue points in Fig.\,\ref{fig:spec}) to 120 and 80 keV for the second and third spectrum (red and black points, respectively).

\section{Conclusion}
\label{sec:concl}

Thanks to the serendipitous \integral\, observations, we have managed to trace the evolution of the hard X-ray flux and the growth of the type-C QPO frequency during the initial stages of the outburst of \swtr. Assuming that the QPO frequency traces the boundary between the disk and the hot flow, we used \integral\ and \srg/\art\ observations to measure the characteristic speed of the inward motion of this boundary during the first two weeks of the outburst. We have also observed switching from type-C to type-B QPO that happened on MJD60205, while the source began the flaring activity.

We have constructed a broad band spectrum of \swtr\ and demonstrated the presence of a hard power-law spectral component up to 400 keV. Additionally, based on the \integral\ data we have traced the evolution of the hard X-ray spectrum during the outburst rise. Finally, we have obtained an upper limit on the hard X-ray flux at the beginning of the optical outburst and estimated the delay of the X-ray outburst with respect to the optical one.

\begin{table}
    \caption{Type-C QPO in \swtr\, (the table in its entirety is available in electronic form only)}
    \label{tab:qpos}
    \centering
    \begin{tabular}{l|c|c|c}
    \hline
         Telescope & MJD, mid-time  & F$_{QPO}$, Hz & F$_{QPO, error}$, Hz  \\
         \hline
         IBIS   & 60181.465   &  0.08 &  0.02\\
         JEM-X  & 60181.817   &  0.14 &  0.01\\
         JEM-X  & 60182.100   &  0.17 &  0.02\\
         JEM-X  & 60182.241   &  0.21 &  0.02\\
         JEM-X  & 60182.351   &  0.24 &  0.03\\
         ART-XC & 60182.884   &  0.37 &  0.01\\
         \hline
    \end{tabular}
\end{table}

\section*{Acknowledgements}

This work is based on observations with the Mikhail Pavlinsky \art\ telescope, hard X-ray instrument on board the \srg\  observatory. The \srg\ observatory was created by Roskosmos  in the interests of the Russian Academy of Sciences represented by its Space Research Institute (IKI) in the framework of the Russian Federal Space Program, with the participation of Germany. 
This work is based on observations with \integral, an ESA project with instruments and the science data centre funded by ESA member states (especially the PI countries: Denmark, France, Germany, Italy, Switzerland, Spain), and Poland, and with the participation of Russia and the USA. 

The work was supported by the RSF grant 19-12-00423. 

\section*{Data availability}

The INTEGRAL, MAXI and Swift data are accessible through the corresponding web pages. At the time of writing, the SRG/ART-XC data and the corresponding data analysis software have a private status. We plan to provide public access to the ART-XC scientific archive in the future. 

\bibliographystyle{mnras}
\bibliography{biblio.bib}

\bsp    
\label{lastpage}
 
\end{document}